# Plasmon-enhanced Brillouin Light Scattering (BLS) spectroscopy for magnetic systems. II. Numerical simulations

Yurii Demydenko[1], Taras Vasiliev[2], Khrystyna O. Levchenko[3], Andrii V. Chumak[3], and Valeri Lozovski[2,4]

[1] V. Lashkariov Institute of Semiconductor Physics National Academy of Sciences of Ukraine, 45 Nauki Avenue, Kyiv, Ukraine
[2] Educational Scientific Institute of High Technologies, Taras Shevchenko National University of Kyiv, 4-g Hlushkova Avenue, Kyiv, 03022, Ukraine
[3] Faculty of Physics, University of Vienna, Boltzmanngasse 5 A, 1090 Vienna, Austria
[4] The Erwin Schrödinger International Institute for Mathematics and Physics, Boltzmanngasse 9, 1090 Vienna, Austria

**Abstract**

Brillouin light scattering (BLS) spectroscopy is a powerful tool for detecting spin waves in magnetic thin films and nanostructures. Despite comprehensive access to spin-wave properties, BLS spectroscopy suffers from the limited wavenumber of detectable spin waves and the typically relatively low sensitivity. In this work, we present the results of numerical simulations based on the recently developed analytical model describing plasmon-enhanced BLS. The effective susceptibility is defined for a single plasmonic nanoparticle in the shape of an ellipsoid of rotation, for the sandwiched plasmonic nanoparticles separated by a dielectric spacer, as well as for the array of plasmonic resonators on the surface of a magnetic film. It is shown that the eccentricity of the metal nanoparticles, which describes their shape, plays a key role in the enhancement of the BLS signal. The optimal conditions for BLS enhancement are numerically defined for gold and silver plasmon systems for photons of different energies. The presented results define the roadmap for the experimental realization of plasmon-enhanced BLS spectroscopy.

**Introduction**

Brillouin light scattering (BLS) spectroscopy [1-7] is a well-established and one of the most powerful tools for the detection of spin waves in the modern field of magnonics, which deals with the processing of data carried by spin waves and their quantum magnons [8-10]. Non-destructive measurements, high dynamic range and ability to measure spin-wave transport with frequency, time, space, wavenumber and phase resolution are the main advantages of BLS spectroscopy. The disadvantages of BLS spectroscopy are the relatively low optical limit of the spin-wave wavenumber detectable by BLS spectroscopy and the relatively low sensitivity, since only a small fraction of the photons from the focused laser are inelastically scattered. Various methods of near-field [11], subdiffraction confinement of the electromagnetic field [12], optically induced Mie resonances in nanoparticles [13], and the use of a dielectric anti-reflection layer [14] have been experimentally tested to extend the wavenumber limit, but have not improved the sensitivity of BLS spectroscopy.

Furthermore, one of the most promising ways to overcome the two mentioned drawbacks of BLS spectroscopy is the utilization of magnon-plasmonic structures [15], which are often exploited to enhance the local field, for example in scanning near-field optical microscopy [16]. A successful example was demonstrated in Ref. [17], where a nanoplasmonic probe, used in a hybrid system of a scanning near-field optical microscope and an atomic force microscope, was able to simultaneously map the optical near-field and the topography of nanostructures. Recently, nanoplasmonic structures have been used in the development of new antiviral methods [18, 19], whose main mechanism relies heavily on the effect of local field enhancement [20].

Therefore, we expect that the implementation of nanoplasmonic structures can be useful for the enhancement of the BLS signal in magnetic systems. The main challenge in engineering plasmon structures for electromagnetic field enhancement is to optimize the morphological and structural configuration of the plasmon systems to maximize the electric field enhancement [21]. In the first part of the work [22], we developed a theoretical analytical model describing the enhancement of the BLS signal as a result of the introduction of the plasmon structure at the surface of the magnetic film. The apparatus is derived from the electrodynamic Green function method within the framework of the effective susceptibility concept [23]. Furthermore, based on the obtained results, a numerical analysis was performed to determine the optimal morphology of nanoplasmonic structures. Two conceptually different cases of plasmon structure modes at the surface were considered: 1) a single plasmon structure, where the plasmon resonances should be considered on the scale of the entire surface of the nanostructure – localized plasmon resonance, and 2) a plasmon structure, where the resonances of the collective excitations should be considered.

Here, a numerical analysis is performed to determine the optimal conditions, such as the morphology of the plasmonic resonators, the metal material of choice, and the photon energy, for the enhancement of the BLS signal using plasmon nanostructures. The presented results define a roadmap for the experimental realization of plasmon-enhanced BLS.

## 1. BLS enhancement by a plasmon structure on the surface of a magnetic film

From the perspective of the magnetic system, the BLS spectroscopy is based on the inelastic scattering of photons with magnons – the quanta of magnetic excitations (spin waves), that takes into account energy and momentum conservation. This interaction leads to a small frequency shift [3] of inelastically scattered photons with respect to elastically scattered ones, which requires high signal intensity and sensitivity. One way to enhance the signal is by utilizing the plasmon structures, that rely on the plasmon resonance for the signal enhancement. In the first part of this work, the general theoretical approach was proposed to study the possibility of enhancing the BLS signal. As a result, the general analytical approach was developed and the analytical expressions, which can further be implemented for numerical calculations, were obtained.

The model of dynamic excitation of the magnetization in magnetic material that creates a small, frequency-dependent change in the susceptibility tensor, was considered in Ref. [2]. In the case of scattering on magnons (spin waves), the change in the susceptibility tensor is caused by the magneto-optical coupling. The magnetization-dependent part of the polarization **P**, induced by the electric field of incident light $\vec{E}_i$, can be written as:

$$P_i = \frac{1}{4\pi}\varepsilon_0 \chi_{ij} E_j, \qquad (1)$$

where $E_j$ is a component of the electric field inside the film, $\chi_{ij}$ is the susceptibility tensor in the first order magneto-optical effects; its relation to the Kerr effect is given by [2], so:

$$\chi_{ij} = \frac{1}{\varepsilon_0}\begin{pmatrix} 0 & KM_3 & -KM_2 \\ -KM_3 & 0 & KM_1 \\ KM_2 & -KM_1 & 0 \end{pmatrix}, \qquad (2)$$

where $M_1$, $M_2$ and $M_3$ are the magnetizations (components) along the $x$, $y$ and $z$ directions, $K$ is the constant of the Kerr effects.

The oscillating electro-dipole (Eq. (1)) can be considered as the secondary source of the scattered (BLS) field. The scattered field is calculated according to:

$$E_i^{(BLS-0)}(\mathbf{R},\bar{\omega}) = -\bar{k}_0^2 \int_{V_F} d\mathbf{R}' G_{ij}^{(32)}(\mathbf{R},\mathbf{R}',\bar{\omega}) P_j(\mathbf{R}',\bar{\omega}), \qquad (3)$$

where $\bar{\omega} = \omega - \Omega$ ($\Omega$ is the frequency shift), $\bar{k}_0 = \bar{\omega}/c$, $G_{ij}^{(32)}(\mathbf{R},\mathbf{R}',\omega)$ is the electrodynamic Green function of the system 'the film at the substrate', where the source of the field is located inside the film and the observation point is in the environment.

A scheme of the standard BLS experiment based on the photon-magnon interaction is demonstrated in Fig. 1. In this scheme, yellow cuboid denoted as '1' is assumed to be Gadolinium Gallium Garnet (GGG) with a dielectric constant $\varepsilon_1$, blue layer '2' is a film of ferrimagnetic dielectric Yttrium Iron Garnet (YIG) with a dielectric constant $\varepsilon_2$, and the environment (air) with a dielectric constant $\varepsilon_3$ is marked with the number '3'. The red arrow depicts the oscillating dipole moment induced due to the scattering of a photon on a magnon (Eq.(1)), which is the source of the electric field of the BLS signal. Black arrows indicate the incident and reflected radiation in the backscattering scheme. Herein, the plasmon structure is located at the surface of a magnetic film and the scattered field, induced by the oscillations of dipole $P$ (Eq.(1)), can be further enhanced by this plasmon structure.

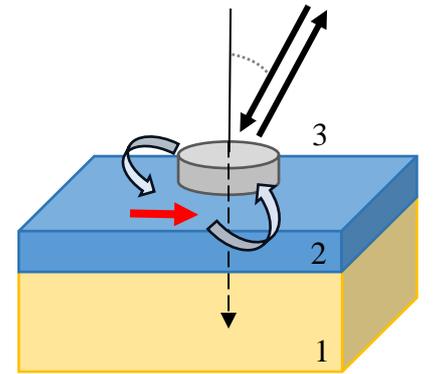

Fig. 1. Scheme of the plasmon structure enhancement of the BLS signal from the magnetic sample. The numbers represent different media with distinct dielectric constants ε: '1' – GGG substrate ($\varepsilon_1$), '2' – YIG film ($\varepsilon_2$), and '3' – the external environment ($\varepsilon_3$).

There are three ways to enhance the BLS signal using the plasmon structure on the surface of the magnetic film. In the first approach, a field induced by the oscillating dipole $P$ at the shifted frequency $\bar{\omega} = \omega - \Omega$ is enhanced by the plasmon structure:

$$E_i^{(BLS-Pl)}(\mathbf{R}, \bar{\omega}) = -\bar{k}_0^2 \int_{V_{Pl}} d\mathbf{R}' G_{ij}^{(33)}(\mathbf{R}, \mathbf{R}', \bar{\omega}) X_{jl}^{(P)}(\mathbf{R}', \bar{\omega}) E_l^{(BLS-0)}(\mathbf{R}', \bar{\omega}) , \qquad (4)$$

where $X_{jl}^{(P)}(\mathbf{R}', \bar{\omega})$ is dimensionless effective susceptibility of plasmon structure [23], $G_{ij}^{(33)}(\mathbf{R}, \mathbf{R}', \omega)$ is the electrodynamic Green function of the system 'the film at the substrate', with the source of the field and the observation point both located in the environment (Fig. 1). The integration in Eq. (4) is over the plasmon structure volume.

The second approach foresees the dipole moment $P$ to be induced by the field at the fundamental frequency, enhanced by the plasmon structure:

$$E_i^{(inc)}(\mathbf{R}, \omega) = -k_0^2 \int_{V_F} d\mathbf{R}' G_{ij}^{(23)}(\mathbf{R}, \mathbf{R}', \omega) X_{jl}^{(p)}(\mathbf{R}', \omega) E_l^{(0)}(\mathbf{R}', \omega), \qquad (5)$$

where $G_{ij}^{(23)}(\mathbf{R}, \mathbf{R}', \omega)$ is the electrodynamic Green function of the system 'the film at the substrate' with the source of the field located in the environment and the observation point – inside the film. Thus, the dipole moment:

$$P_i(\mathbf{R}, \bar{\omega}) = -\frac{k_0^2}{4\pi} \chi_{ij} \int_{V_F} d\mathbf{R}' G_{jk}^{(23)}(\mathbf{R}, \mathbf{R}', \omega) X_{kl}^{(p)}(\mathbf{R}', \omega) E_l^{(0)}(\mathbf{R}', \omega) \qquad (6)$$

induces the field on the detector:

$$E_i^{(BLS-Pl)}(\mathbf{R},\overline{\omega}) = \overline{k}_0^2 \frac{k_0^2}{4\pi} \int_{V_F} d\mathbf{R}' G_{ij}^{(32)}(\mathbf{R},\mathbf{R}',\overline{\omega})\chi_{jm}$$
$$\times \int_{V_F} d\mathbf{R}'' G_{mn}^{(23)}(\mathbf{R}',\mathbf{R}'',\omega) X_{nl}^{(p)}(\mathbf{R}'',\omega) E_l^{(0)}(\mathbf{R}'',\omega) .$$
(7)

Finally, the third way is a combination of the first two approaches: the dipole moment is induced by the enhanced field (the enhancement is at the fundamental frequency) and the field already induced by this oscillating moment (at the shifted frequency) is enhanced by the plasmon structure:

$$E_i^{(BLS-Pl)}(\mathbf{R},\overline{\omega}) = \overline{k}_0^2 \frac{k_0^2}{4\pi} \int_{V_p} d\mathbf{R}' G_{ij}^{(33)}(\mathbf{R},\mathbf{R}',\overline{\omega}) X_{jk}^{(p)}(\mathbf{R}',\overline{\omega}) \int_{V_F} d\mathbf{R}''' G_{kp}^{(32)}(\mathbf{R}',\mathbf{R}''',\overline{\omega})\chi_{pm}$$
$$\times \int_{V_F} d\mathbf{R}'' G_{mn}^{(23)}(\mathbf{R}''',\mathbf{R}'',\omega) X_{nl}^{(p)}(\mathbf{R}'',\omega) E_l^{(0)}(\mathbf{R}'',\omega) .$$
(8)

Obviously, the enhancement of the BLS signal can be provided under the condition of 'plasmon resonance' in the plasmon system, when the $X_{jl}^{(P)}(\mathbf{R}',\omega)$ maximum is reached. Analysis of the behavior of the effective susceptibilities of plasmon structures is the main challenge discussed in this part of the work.

Earlier, we have mentioned that there are two methods to build the plasmon structures in which the arising resonance conditions can be used for enhancement of the BLS signal:

*i.* The first approach is to use the single plasmon structures. For example, the BLS signal can be observed when the magnet substrate is a thin strip of a width about the wavelength of a magnon [24]. In this case, the plasmonic structure will consist of some (small) number of nanoobjects on the surface of the YIG film. These can be either single metal nanoparticles or complex nanoparticles, e.g., two metal nanoparticles separated by a dielectric spacer (Fig. 2). In analogue to Fig. 1, numbers '1', '2', '3' represent the GGG substrate, YIG film and external media (air) respectively. Number '4' is an example of the aforementioned complex plasmonic structure at the surface of a magnetic film. Presented nanoplasmonic structure can enhance the BLS by plasmon resonance.

*ii.* The second approach is to cover the surface of the magnetic film with the submonolayer of the nanoparticles, which can be assumed as an infinitive in the surface plane. The condition of the BLS signal enhancement, in this case, should be the excitation of collective mode characterized by a wave vector (for example the surface plasmon resonance located at the submonolayer of metal nanoparticles). Thus, the resonance will have the character of configurational resonance [25], with its conditions depending on the structure of a single nanoobject and the concentrations of the particles.

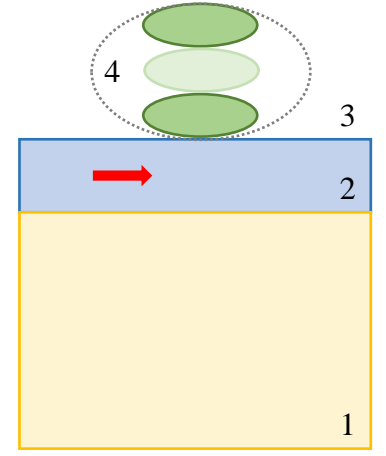

Fig. 2. Sketch of a single complex plasmon nano-structure at the magnetic film surface. The numbers represent different media: '1' – GGG substrate, '2' – YIG film, '3' – the external environment, '4' – exemplary nanoplasmonic structure.

## 2. Effective susceptibility of plasmon structures

To establish the configurations of plasmon structures, one should know their effective susceptibilities. Let us assume a single plasmon structure on the top of a magnetic film, which could potentially provide a BLS signal enhancement (Fig. 2). Consider, for clarity and simplicity, the nanostructure consisting of three ellipsoids located one above the other. In practice, the results obtained in this work can be qualitatively projected onto the disks, which can be easily fabricated using modern nanostructuring techniques. However, the precise qualitative analysis for the disks would require more specialized numerical studies. The effective susceptibility of such a structure, according to the method developed in Ref. [23], was calculated in the first part of this work as:

$$X_{ji}^{(S)}(\mathbf{R}^{(\alpha)},\omega) = \left\{ \left[\chi_{ji}^{(\alpha)}(\omega)\right]^{-1} + k_0^2 \int_{V_\alpha} d\mathbf{R}' G_{ij}^{(33)}(\mathbf{R}',\mathbf{R}^{(\alpha)},\omega) + \right.$$
$$\left. + k_0^2 \sum_{\beta \neq \alpha} \int_{V_\beta} d\mathbf{R}^{(\beta)} G_{ij}^{(33)}(\mathbf{R}^{(\beta)},\mathbf{R}^{(\alpha)},\omega) \right\}^{-1}, \quad \alpha,\beta = 1,2,3 , \quad (9)$$

where the integrand in the second term in a bracket contains the indirect part of the Green function [26] and $\chi_{ij}^{(\alpha)}(\omega)$, $\alpha = 1, 2, 3$ are the susceptibilities of single nanoparticles forming the plasmon structure (Fig. 2).

These susceptibilities can be calculated as the linear response of the ellipsoidal particle located inside the homogeneous isotropic medium (air, for example), which is characterized by the dielectric constant $\varepsilon_3 = 1$ and the depolarization factors $m_i$:

$$\chi_{ij}^{(\alpha)}(\omega) = \chi^{(\alpha)}(\omega) \frac{1}{1 + 4\pi \chi^{(\alpha)}(\omega) m_i} \delta_{ij}, \qquad (10)$$

The components of the depolarization factor depend on the geometry, namely:

– for prolate ellipsoids with rotation ($h_z > h_x = h_y$), where the $h_i$ are semiaxes of ellipsoids and $h_z$ rotation axis around OZ is normal to the film surface

$$m_z = \frac{1 - \zeta^2}{\zeta^3} \left( \frac{1}{2} \ln \frac{1+\zeta}{1-\zeta} - \zeta \right), \quad m_x = m_y = \frac{1}{2}(1 - m_z), \quad \zeta = \sqrt{1 - h_x^2 / h_z^2}, \qquad (11)$$

– for oblate ellipsoids with rotation ($h_x = h_y > h_z$):

$$m_z = \frac{1 + \xi^2}{\xi^3} \left( \xi - \arctan \xi \right), \quad m_x = m_y = \frac{1}{2}(1 - m_z), \quad \xi = \sqrt{h_x^2 / h_z^2 - 1}, \qquad (12)$$

where $\chi^{(\alpha)}(\omega)$ is a dielectric response of the material from which the 'α-th' nanoparticle is fabricated. Integrations in Eq. (5) are over the volumes of nanoparticles 1, 2 and 3, $G_{ij}^{(33)}(\mathbf{R}, \mathbf{R}', \omega)$ is the Green function of the film at the substrate, in which the source of the field and the observation point are located in the environment '3' (note, the Green function was calculated in so-called '**k**-$z$ representation') in Ref. [27]).

Now, let us look at the plasmon system consisting of a magnetic film covered with nanoparticles of a concentration $n$ (Fig. 1). If the nanoparticles at the sub-monolayer cover of the surface are of metallic origin, one can expect them to form ellipsoids of rotation, and the effective susceptibility (the linear response to the external relatively to the particle field) of a single nanoparticle, in that case, would be [28]:

$$\tilde{\chi}_{ii} = \varepsilon_3 V_p \frac{(\varepsilon_p - \varepsilon_3)}{\varepsilon_3 + (\varepsilon_p - \varepsilon_3)m_i} L_\parallel, \quad i = x, y, \tag{13}$$

$$\tilde{\chi}_{zz} = \varepsilon_3 V_p \frac{(\varepsilon_p - \varepsilon_3)}{\varepsilon_3 + (\varepsilon_p - \varepsilon_3)m_i} L_\perp, \tag{14}$$

where

$$L_\parallel = \left[1 + \frac{(\varepsilon_3 - \varepsilon_2)(\varepsilon_p - \varepsilon_3)}{3(\varepsilon_3 + \varepsilon_2)(\varepsilon_3 + (\varepsilon_p - \varepsilon_3)m_i)} \vartheta \right]^{-1}, \tag{15}$$

and

$$L_\perp = \left[1 + \frac{(\varepsilon_3 - \varepsilon_2)(\varepsilon_p - \varepsilon_3)}{3(\varepsilon_3 + \varepsilon_2)(\varepsilon_3 + (\varepsilon_p - \varepsilon_3)m_i)} 2\vartheta \right]^{-1} \tag{16}$$

are the parameters describing the shape of the particles, with depolarization factors according to Eq. (11) and Eq.(12). The value $\varepsilon_2$ is the dielectric constant (in the optical range of wavelengths) of the magnetic film's material, value $\varepsilon_3$ is the dielectric constant of the environment, value $\varepsilon_p$ is the dielectric constant of the metal from which the nanoparticle is fabricated.

Parameter $\vartheta$, determined by the particle dimension, is written as follows:

$$\vartheta = h_x h_y h_z (2z_p)^{-3}, \tag{17}$$

where $z_p$ is the distance from the particle center to the surface. If this distance equals the length of the semi-axis along OZ axis of the coordinate system $z_p = h_z$, then:

$$\vartheta = (h_x / h_z)^2 \cdot (1/8) \tag{18}$$

According to work [23], the effective susceptibility of a submonolayer cover of the nanoparticles, which can be used for the calculation of the BLS signal, is:

$$X_{ij}(\bar{\omega}, k, d) = \left[ \left( \tilde{\chi}_{\substack{ii \\ zz}}(\omega) \right)^{-1} \delta_{\substack{ji \\ zz}} + n k_0^2 G_{ji}^{(k)}(k, d, \bar{\omega}) \right]^{-1}, \tag{19}$$

where $n$ is a concentration of nanoparticles at the surface of magnetic film, and $G_{ji}^{(k)}(k,d,\bar{\omega})$ – the electromagnetic Green function of the system 'film at the substrate' [27]. Note, that Eq. (19) for the effective susceptibility of the submonolayer cover of the nanoparticles is restricted by the maximum concentration so that the additional nanoparticle can be placed between the two neighboring particles at the maximum concentration. In other words, the projections of the particles onto the surface of a magnetic film should be such that one can place the other projection between the two others (Fig. 3). This restriction means that the nanoparticles polarize as the particles with a certain volume and a shape, but they interact as the point-like dipoles [29]. Precisely in the frame of such considerations, Eq. (19) was obtained in Ref. [23].

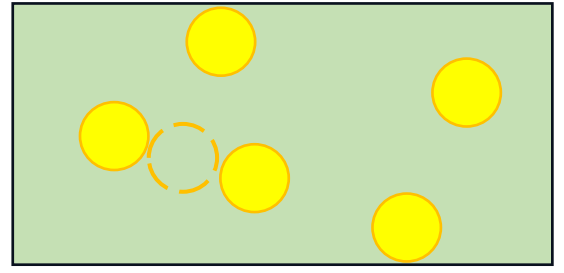

Fig. 3. Visualization of the maximum concentration of the particles – only one additional nanoparticle can be placed between the two neighboring ones.

3. **Numerical calculations of BLS enhancing assisted by plasmon structure**

As mentioned above, the enhancement of the BLS signal assisted by the plasmon structure can be observed under a condition of 'plasmon resonance'. The 'pure' plasmon resonance condition means formally that the pole part of the effective susceptibilities (Eq. (9) and Eq. (19)) must go to zero. Hence, the condition for enhancing the BLS signal by plasmon structure, when the integrals in Eqs.(4-8) reach their maximum values, is closely connected to this 'plasmon resonance'. Therefore, it can be observed when the effective susceptibilities reach maximum, or the denominators of the susceptibilities reach minimum. Hence, one has to solve the optimization problem to find out the parameters of plasmon structures at which the signal enhancement can be observed:

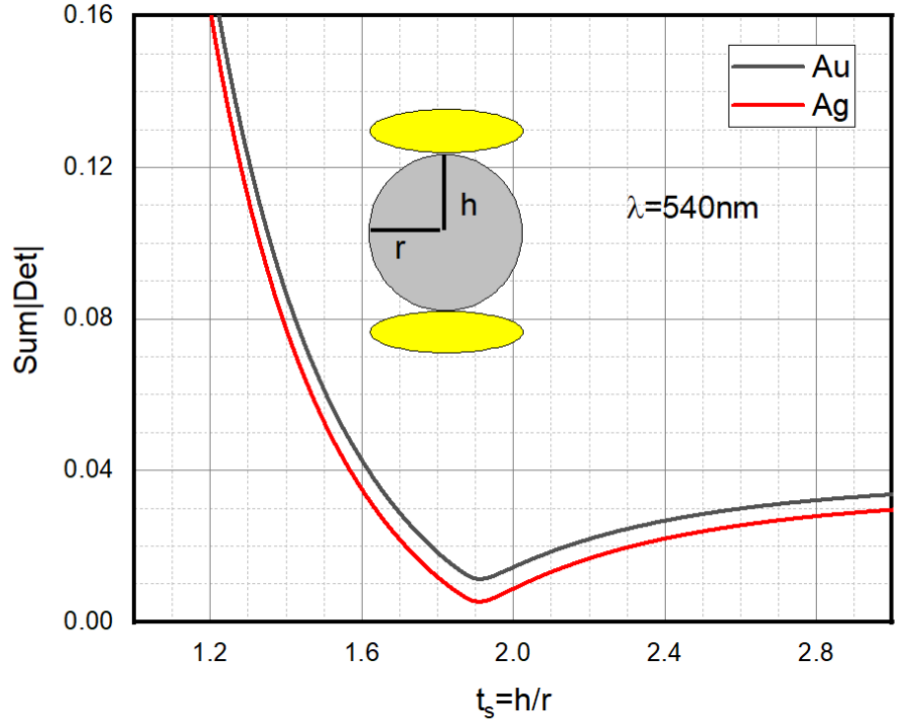

Fig. 4. The behavior of a denominator of the effective susceptibility of plasmon structure if the eccentricity of metal nanoparticles is equal to 0.25.

$$M\det = \min\Biggl[\Bigl|\mathrm{Re}\,\mathrm{Det}\Bigl\{\bigl[\chi_{ji}^{(\alpha)}(\omega)\bigr]^{-1} + k_0^2\int_{V_\alpha} d\mathbf{R}' G_{ij}^{(3-3)}(\mathbf{R}',\mathbf{R}^{(\alpha)},\omega) +$$

$$+ k_0^2 \sum_{\beta\neq\alpha}\int_{V_\beta} d\mathbf{R}^{(\beta)} G_{ij}^{(3-3)}(\mathbf{R}^{(\beta)},\mathbf{R}^{(\alpha)},\omega)\Bigr\}\Bigr|^2 +$$

$$+\Bigl|\mathrm{Im}\,\mathrm{Det}\Bigl\{\bigl[\chi_{ji}^{(\alpha)}(\omega)\bigr]^{-1} + k_0^2\int_{V_\alpha} d\mathbf{R}' G_{ij}^{(3-3)}(\mathbf{R}',\mathbf{R}^{(\alpha)},\omega)$$

$$+ k_0^2 \sum_{\beta\neq\alpha}\int_{V_\beta} d\mathbf{R}^{(\beta)} G_{ij}^{(3-3)}(\mathbf{R}^{(\beta)},\mathbf{R}^{(\alpha)},\omega)\Bigr\}\Bigr|^2\Biggr]^{1/2}, \quad \alpha,\beta=1,2,3, \qquad (20)$$

for the first type of plasmon structure, and

$$M\det = \min\left\{\left|\text{Re}\,\text{Det}\left[\left(\tilde{\chi}_{ii\atop zz}(\omega)\right)^{-1}\delta_{ji\atop zz} + nk_0^2 G_{ji}^{(k)}(k,d,\overline{\omega})\right]\right|^2\right.$$

$$\left. + \left|\text{Im}\,\text{Det}\left[\left(\tilde{\chi}_{ii\atop zz}(\omega)\right)^{-1}\delta_{ji\atop zz} + nk_0^2 G_{ji}^{(k)}(k,d,\overline{\omega})\right]\right|^2\right\}^{1/2}$$

(21)

for the second type of plasmon structure. Both of these conditions are numerically analyzed below.

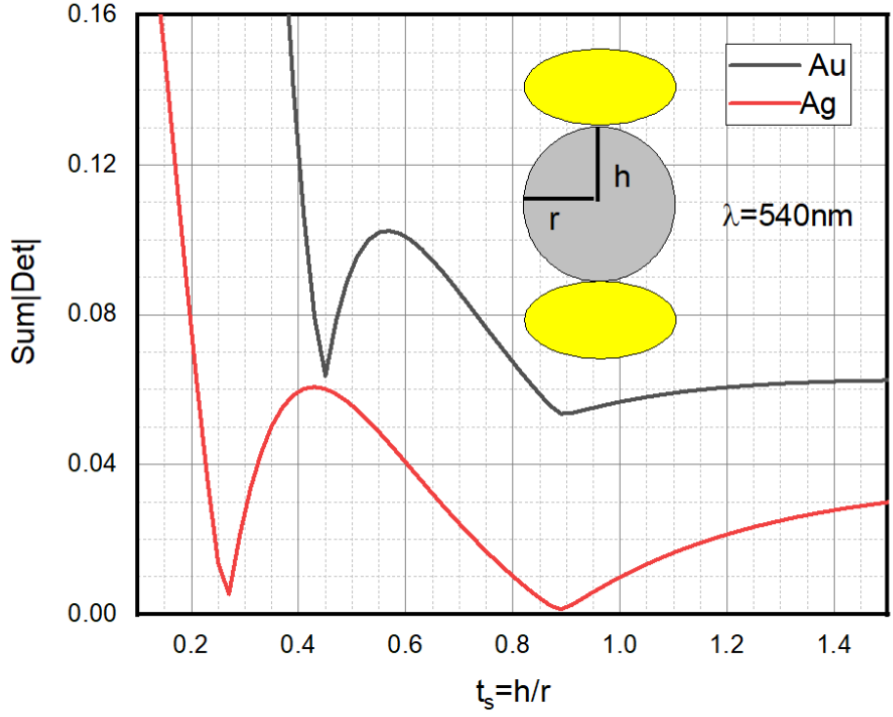

Fig. 5. The behavior of a denominator of the effective susceptibility of plasmon structure if the eccentricity of metal nanoparticles equals to 0.5.

### 3.1. BLS enhancement assisted by a single plasmon structure

Firstly, we consider the possibility of the BLS signal enhancement by single complex nanoparticles forming a plasmon structure. Numerical calculations of Eq. (20) provide us following results (Figs. 4-6). According to Eq. (20), the plasmon structure capable of enhancing the BLS signal consists of the two metal nanoparticles, shaped as ellipsoids of rotation with radii in the plane of the film equal to 10 nm, and separated by the dielectric spacer, shaped as an ellipsoid of rotation. Results of the calculations when the eccentricity of the metal nanoparticles is 0.25 are shown in Fig. 4.

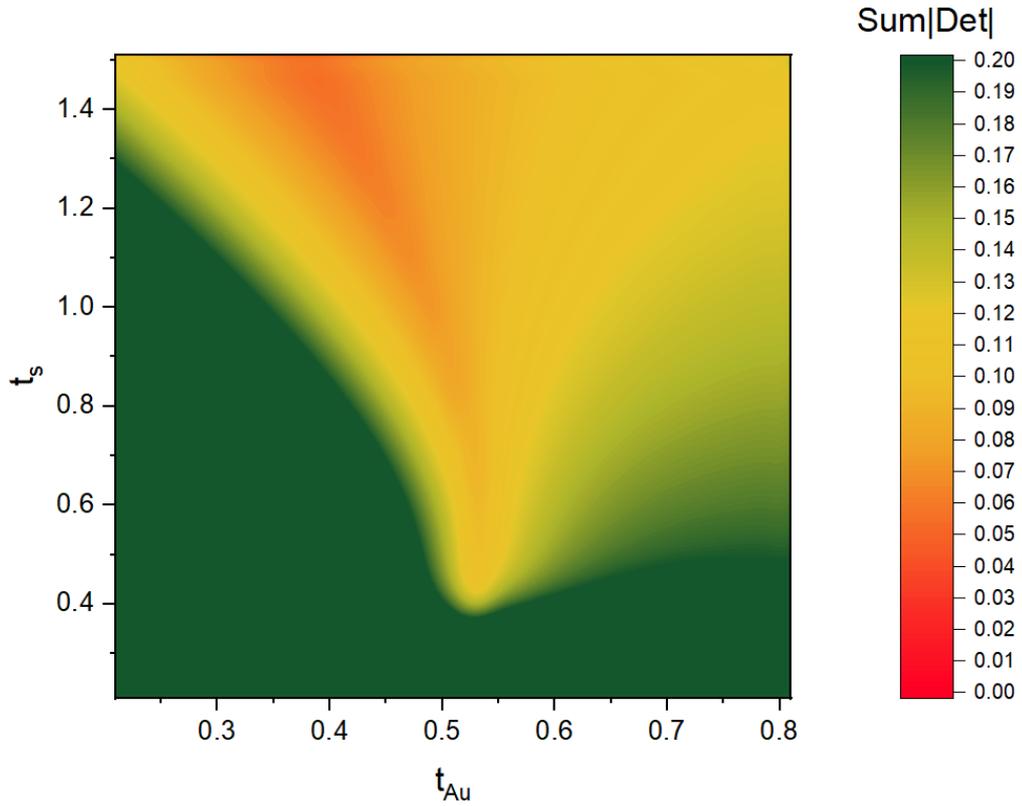

Fig. 6. The optimization map of the morphology of the golden plasmon structure consisting of two nanoparticles shaped as ellipsoids of rotation with the ellipsoidal dielectric spacer. $t_s$ and $t_{Ag}$ are the eccentricities of the spacer and metal nanoparticles, respectively.

Figure 4 demonstrates how the enhancement conditions in the case of single complex nanoparticles with an eccentricity of 0.25 are fulfilled when the eccentricity of a spacer is about 1.95. Thus, the spacer should be prolate ellipsoid. Both gold and silver nanoparticles, for example, can be used for plasmon structures to enhance the BLS signal.

The results of similar calculations for the plasmon structure, if the eccentricity of metal nanoparticles is 0.5, are shown in Fig. 5. Here, the shape of the nanoparticle of a spacer is oblate, with the eccentricity about 0.93. Note, that in this case, silver nanoparticles provide better conditions for BLS signal enhancement. Moreover, two minima of the pole part of the effective susceptibility are observed (Fig. 5), which is beneficial for the optimization problem solution.

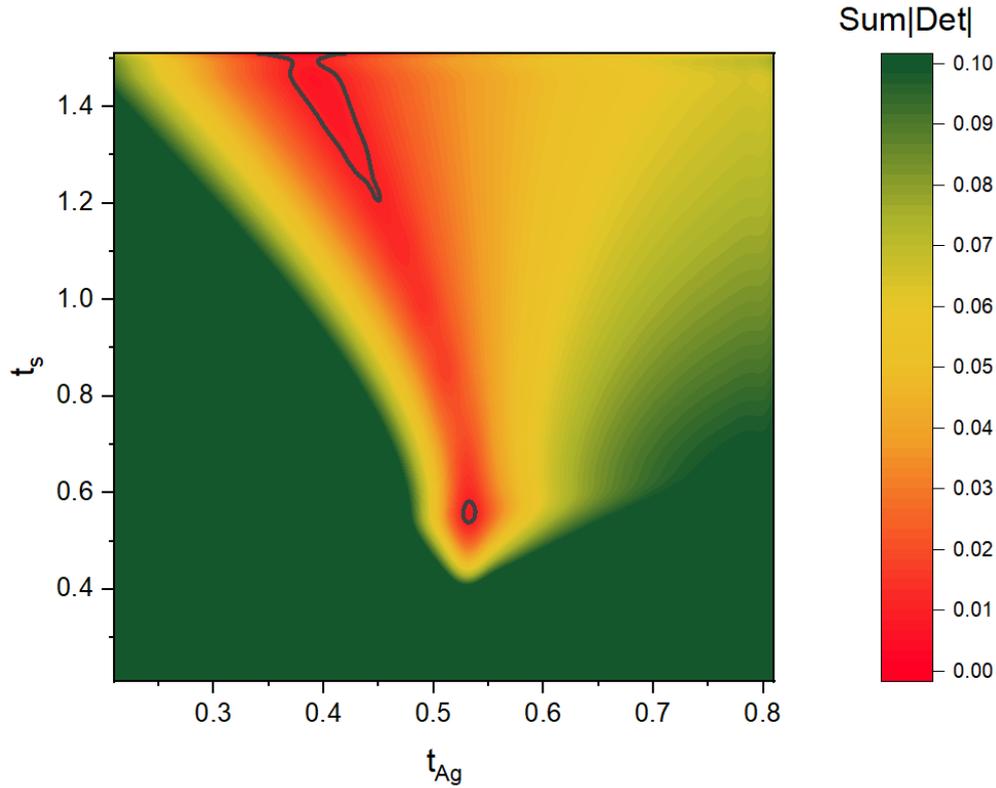

Fig. 7. The optimization map of the morphology of the silver plasmon structure consisting of two nanoparticles shaped as ellipsoids of rotation with the ellipsoidal dielectric spacer. $t_s$ and $t_{Au}$ are the eccentricities of spacer and metal nanoparticles, respectively.

The results of calculations focused on solving a problem of the optimization of morphological parameters of the plasmon structure are depicted in Fig. 6 and Fig. 7. The outcomes are presented as a 3D graph in the green-red scale. As shown in Fig. 6, the gold plasmon structure can demonstrate the enhancement effect when the eccentricity of golden nanoparticles is about 0.55-0.5 and the eccentricity of the spacer is about 0.43-0.63. Moreover, the BLS enhancement effect is also observed if the eccentricity of golden nanoparticles is in the range of 0.35-0.5 and when the eccentricity of the spacer lies between 0.96-1.5.

The results of similar calculations for plasmon structures fabricated from silver are shown in Fig. 7. Contrary to the golden plasmon structure, in this case, there is rather a narrow parameter gap that facilitates the BLS signal enhancement. Moreover, similar optimization calculations strongly restrict the range of parameters of the plasmon structures for further experiment planning. Note, in Fig .6 and Fig. 7 the domains where $M$det < 0.001 are marked by a dark line.

The question arises whether it is possible to obtain BLS signal enhancement using a simple nanoplasmonic structure consisting of a single particle on the surface. To answer this question, one has to consider a single metal nanoparticle at the surface of a magnetic film (Fig. 8). In this case, the optimization problem is simplified, and is reduced to the analysis of the expression:

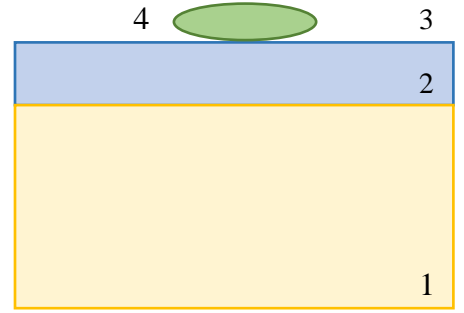

Fig. 8. Sketch of a single metal nanoparticle at the surface of the magnetic film. The numbers represent different media: '1' – GGG substrate, '2' – YIG film, '3' – the external environment, '4' – exemplary nanoparticle plasmonic structure.

$$\min = M \det = \left( \left| \operatorname{Re} \det \left\{ \left[ \chi_{ji}^{(1)}(\omega) \right]^{-1} + M_{ij}^{(1-1)}(\mathbf{R}^{(1)}, \omega) \right\} \right|^2 \right.$$
$$\left. + \left| \operatorname{Im} \det \left\{ \left[ \chi_{ji}^{(1)}(\omega) \right]^{-1} + M_{ij}^{(1-1)}(\mathbf{R}^{(1)}, \omega) \right\} \right|^2 \right)^{1/2}, \qquad (22)$$

with self-energy part

$$M_{ij}^{(1-1)}(\mathbf{R}^{(1)}, \omega) = k_0^2 \int_{V_1} d\mathbf{R}' G_{ij}^{(B)}(\mathbf{R}', \mathbf{R}^{(1)}, \omega) . \qquad (23)$$

The numerical calculations were performed for the golden and silver particles with an in-plane diameter of 20 nm. The results are presented as a map of frequencies and eccentricities in the form of a green-red scale (Fig. 9 and Fig.10). From the figures one can estimate that in the case of simple plasmon structure, the conditions of the BLS enhancement (Eq.(22)) are fulfilled with the accuracy $M \det < 0.001$. Thus, the plasmon nanostructures composed either of three nanoparticles (two metal particles separated by a spacer) or of a single metal nanoparticle, can effectively enhance the BLS signal. It should be noted that in the case of a single nanoparticle, the conditions of plasmon resonance are not fulfilled. However, a sharp decrease in the denominator in the effective susceptibility will still lead to an increase in the integral in the right part of Eq. (8), i.e., to the enhancement of the BLS signal.

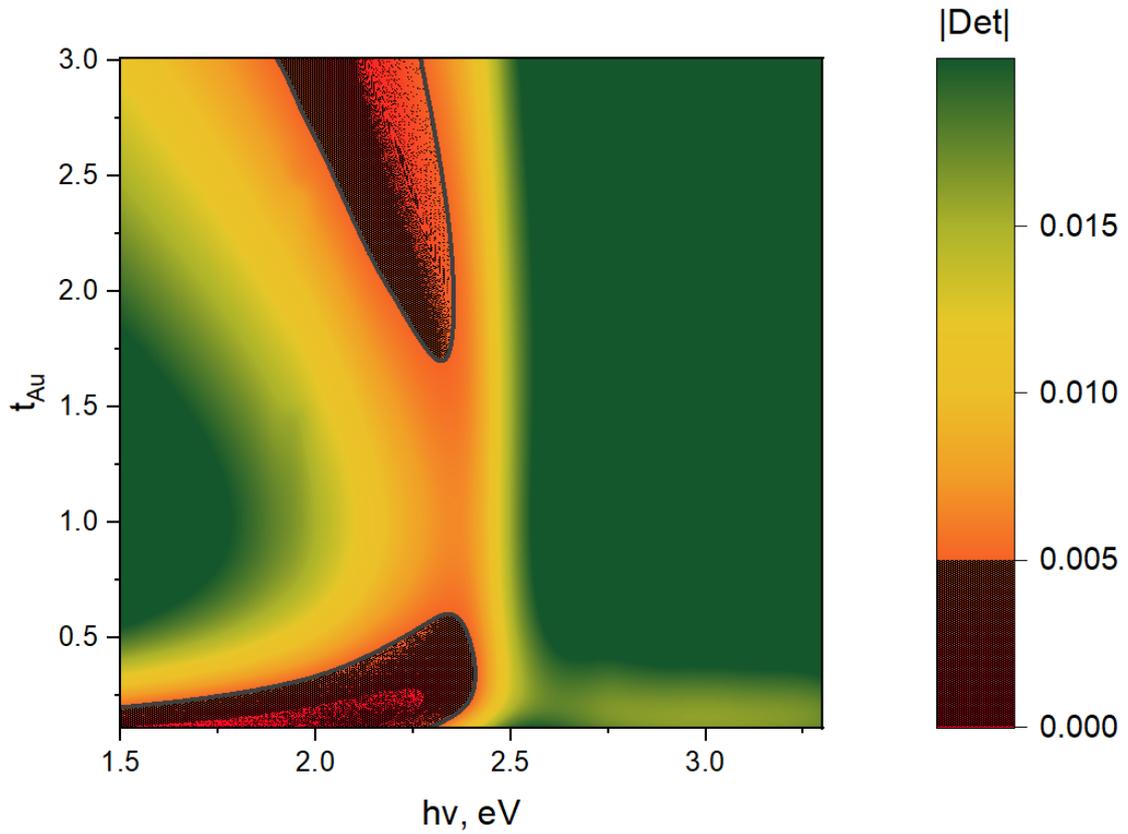

Fig. 9. The optimization map of values $M\text{det}$ calculated for a single golden nanoparticle. The domains $M\det < 0.001$ are marked by the dashed lines.

Similar calculations were performed for the silver nanoparticles. The results are shown in Fig. 10. The branching of the region, which corresponds to flattened particles, looks the best in terms of possible practical application when the conditions of the fabrication of the nanoplasmonic structures are met. This will be discussed in more detail further.

In practice, the entire visible range and the shape of the particle can be selected so that the resonant enhancement of the local field is observed. At the same time, the longer the wavelength of the incidence light, the flatter the particle should be.

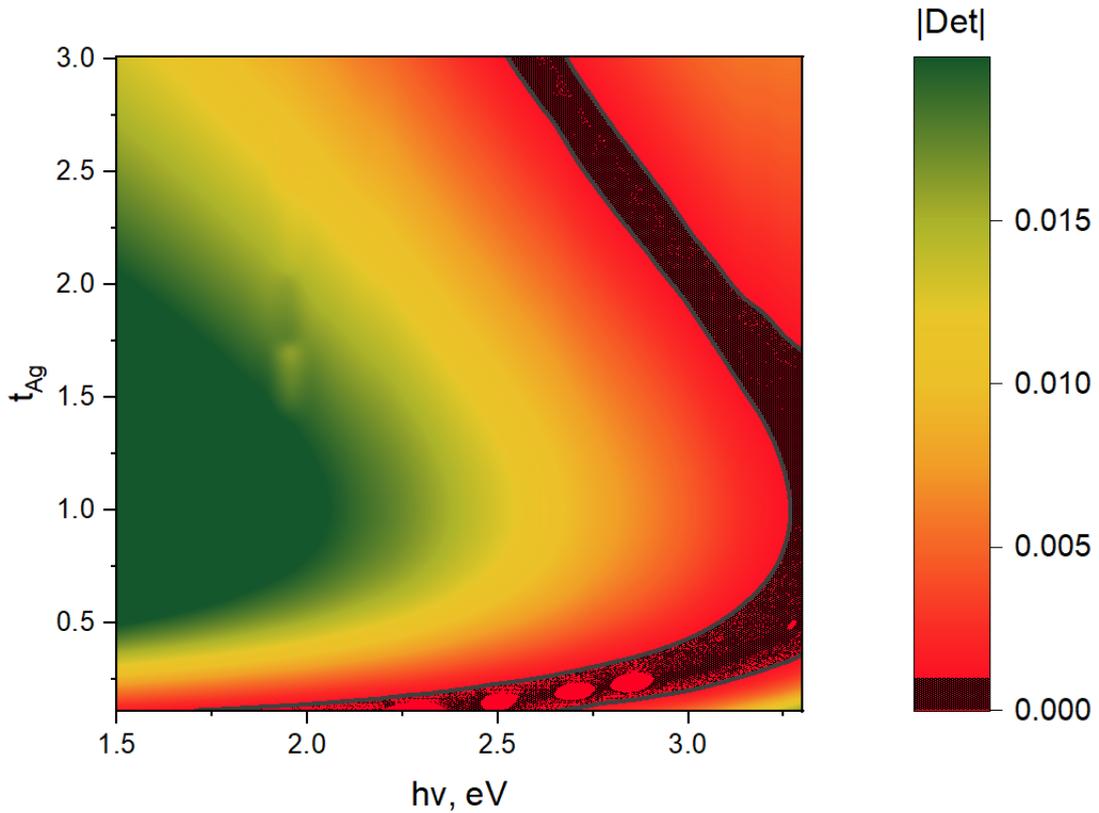

Fig. 10. The optimization map of values $M\text{det}$ calculated for a single silver nanoparticle. The domains $M\text{det} < 0.001$ are marked by dashed lines

## 3.2. BLS signal enhancement assisted by plasmon structure in the form of a magnetic film surface covered with nanoparticles

Consider now the second kind of plasmon structure – magnetic film covered by metal nanoparticles. It is clear that the enhancement by a multiparticle system can be more effective than by a single nanoparticle since in the former a mode localized on the entire multiparticle system can be used. However, at the same time, an additional problem arises related to the formation of the almost 'surface' wave. On one hand, the BLS signal must be a wave belonging to the radiation region – it must propagate from the surface of the magnetic film to the detector, and on the other hand, the surface wave must be an evanescent wave, that is, a wave that is localized near the surface and does not propagate in the direction perpendicular to the surface. Further in the work, we estimate if these conditions can be fulfilled.

It can be seen from Eq. (8) that the enhancement of the BLS field can be ensured both by the 'resonant' behavior of the effective susceptibility of the plasmon nanosystem at the frequency of the incident radiation ω and at the shifted frequency ω - Ω (ω + Ω). As it was noted in the first part of the work, since the propagation constant $\sqrt{\overline{k}_0^2 - \left(k_\parallel^{BLS}\right)^2}$, $\overline{k}_0 = (\omega - \Omega)/c$ under the conditions of backscattering, when $k_\parallel^{BLS} = -k_0 \sin\vartheta$, must be real. It can be understood that from the condition $\left(\frac{\omega-\Omega}{c}\right)^2 - \left(\frac{\omega}{c}\sin\vartheta\right)^2 \geq 0$, one arrives to $\cos^2\vartheta - 2\frac{\Omega}{\omega} + \frac{\Omega^2}{\omega^2} \geq 0$. Hence, there is a limitation on the incidence angle of the radiation $\vartheta$: $\cos\vartheta \geq \sqrt{2\frac{\Omega}{\omega} - \frac{\Omega^2}{\omega^2}}$. Thus, it is not possible to use angles of incidence in a narrow region near $\vartheta = \pi/2$.

Then, for the root expression one has $\left(\frac{\omega-\Omega}{c}\right)^2 - \left(\frac{\omega}{c}\sin\vartheta\right)^2$, yet a condition for the surface wave existence (the wave must be evanescent one) is $\left(\frac{\omega}{c}\right)^2 - \left(\frac{\omega}{c}\sin\vartheta\right)^2 < 0$. But in the case of backscattering wave $\left(\frac{\omega}{c}\right)^2 - \left(\frac{\omega}{c}\sin\vartheta\right)^2 = \left(\frac{\omega}{c}\cos\vartheta\right)^2 > 0$. That is, the condition for the excitation of the evanescent wave cannot be fulfilled without additional effort. Thus, BLS enhancement cannot be realized under the conditions of the surface plasmon resonance in the layer of the nanoparticles and the enhancement will not be effective enough.

Considering these circumstances, we carried out numerical calculations of the parameters of the nanosystem for which we can expect the BLS signal enhancement. It is established that the BLS improvement should depend on the polarization of incident light. In previous calculations, we have used a model consisting of a magnetic (yttrium iron garnet, YIG) film covered by a golden ellipsoid with a diameter (projection onto the film's plane) $d$ = 20 nm. The maximum concentration at which one can use the quasi-point-dipole approximation considering a particle-particle interaction [29] is $6,25 \cdot 10^{10}$ cm$^{-2}$. The results of the optimization calculations considering nonpolarized incident light are shown in Fig. 11. Comparing the results shown in Fig. 9 and Fig. 11, one can see, that for the light of the wavelength ranging ~ 420-600 nm, the images of optimization are very similar. This can be understood if to consider that in both cases the resonances of single particles play the main role. Only for high-energy photons the interparticle interactions, which form the collective mode, start to gain importance. Similar calculations, but individually for the incident light of s- and p-polarization, are shown in Fig. 12. One can see that the value of $M$det for the polarized light is larger than for nonpolarized light and larger in comparison to the $M$det calculated for a single nanoobject. This explains employment of a weaker reinforcement condition $M\det < 0.005$ in Fig. 11 instead of $M\det < 0.001$, as it was used for a single nanostructure and for silver submonolayer.

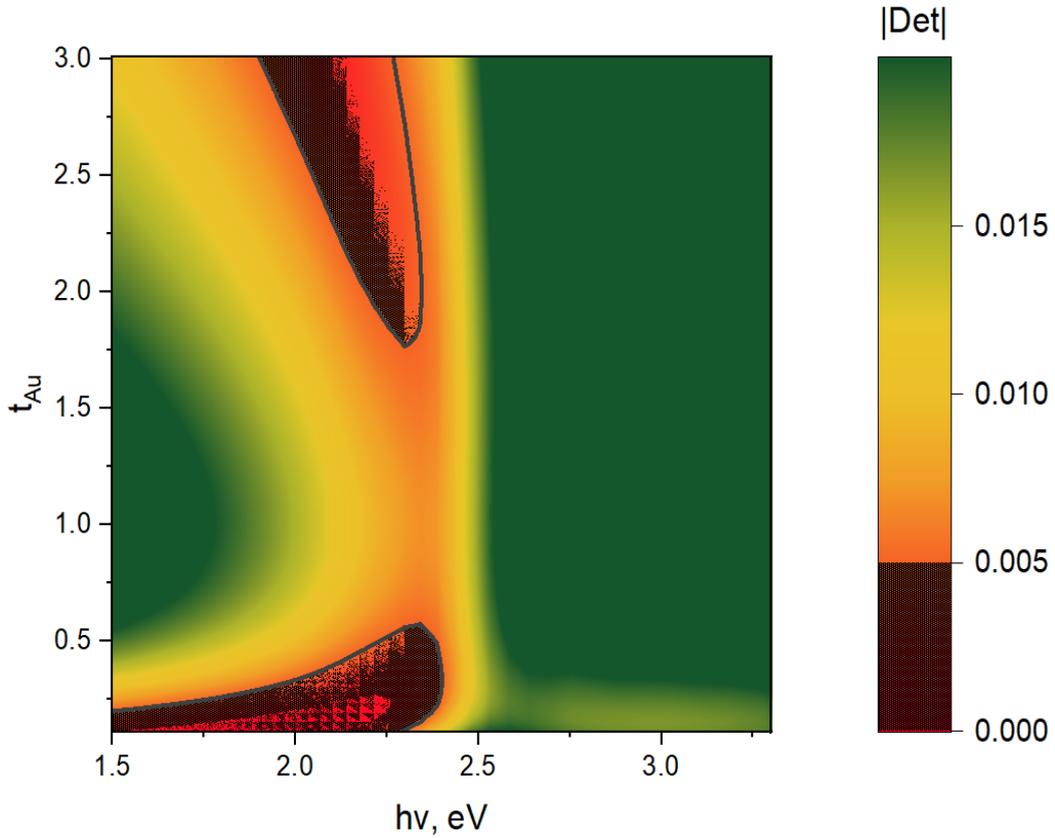

Fig. 11. The optimization map of values $M$det calculated for a submonolayer of the golden nanoparticles of diameter 20 nm seeded onto a YIG film of 100 nm thickness. The angle of the incident light was considered to be 10°. The domains $M\det < 0.001$ are marked by the dashed lines.

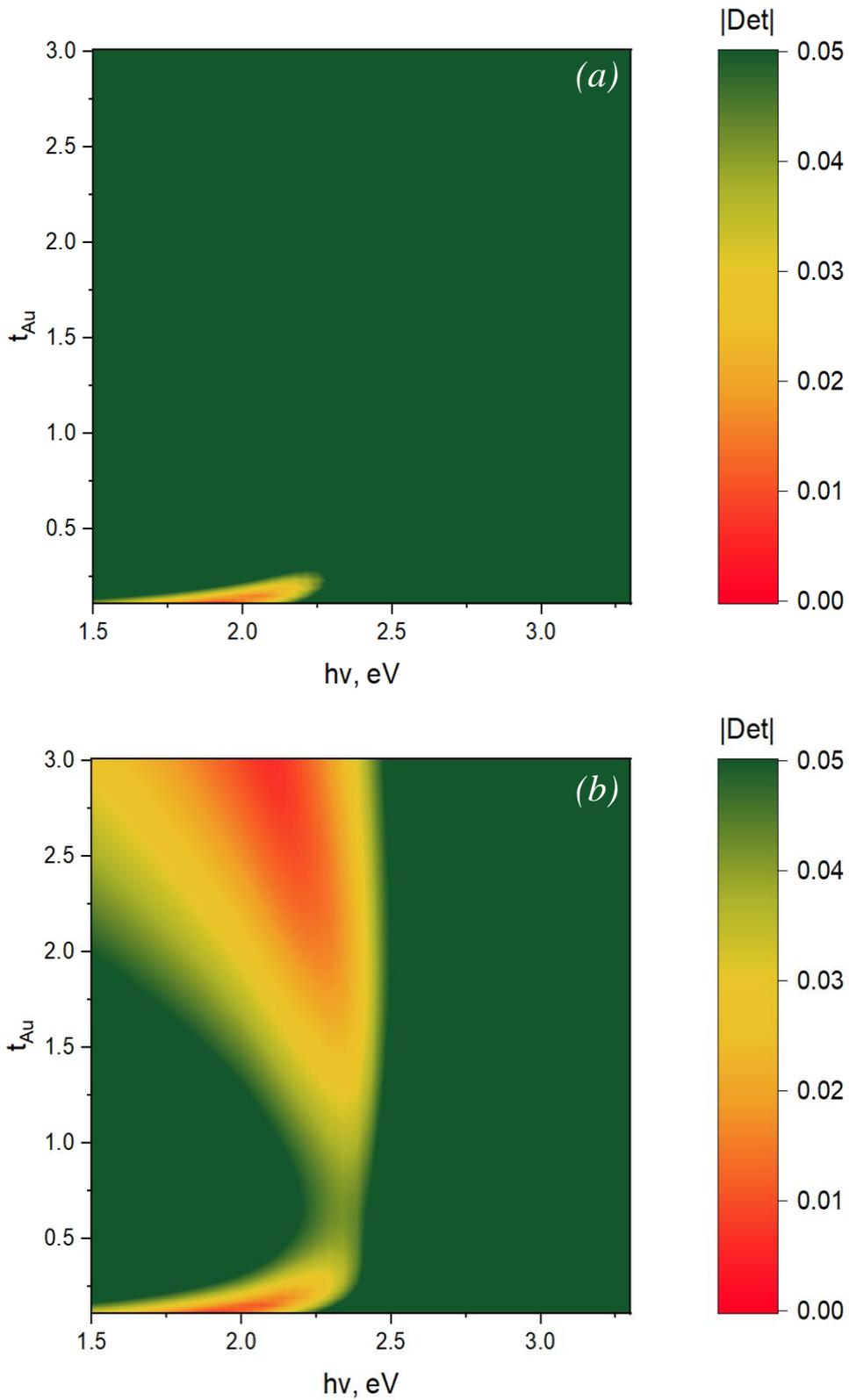

Fig. 12. The optimization map of values *M*det calculated for a submonolayer of golden nanoparticles of diameter 20 nm seeded onto a YIG film of 100 nm thickness. The angle of the incident light was set considered to be 10°.

The optimization map of values *M*det calculated for a submonolayer of the silver nanoparticles of diameter 20 nm seeded onto a YIG film of 100 nm thickness is shown in Fig. 13. The angle of the incident nonpolarized light is considered to be 10°. From the figures, one can deduce that in the case of nonpolarized light, both the optimization images for a single silver nanoparticle (Fig. 10) and for a full surface covered by the silver nanoparticles (Fig. 13) are very similar. This suggests that single-particle resonances play a key role here. As it was mentioned above, the reason for this behavior lies in the nonresonant excitation of the waves along the nanoparticle plane.

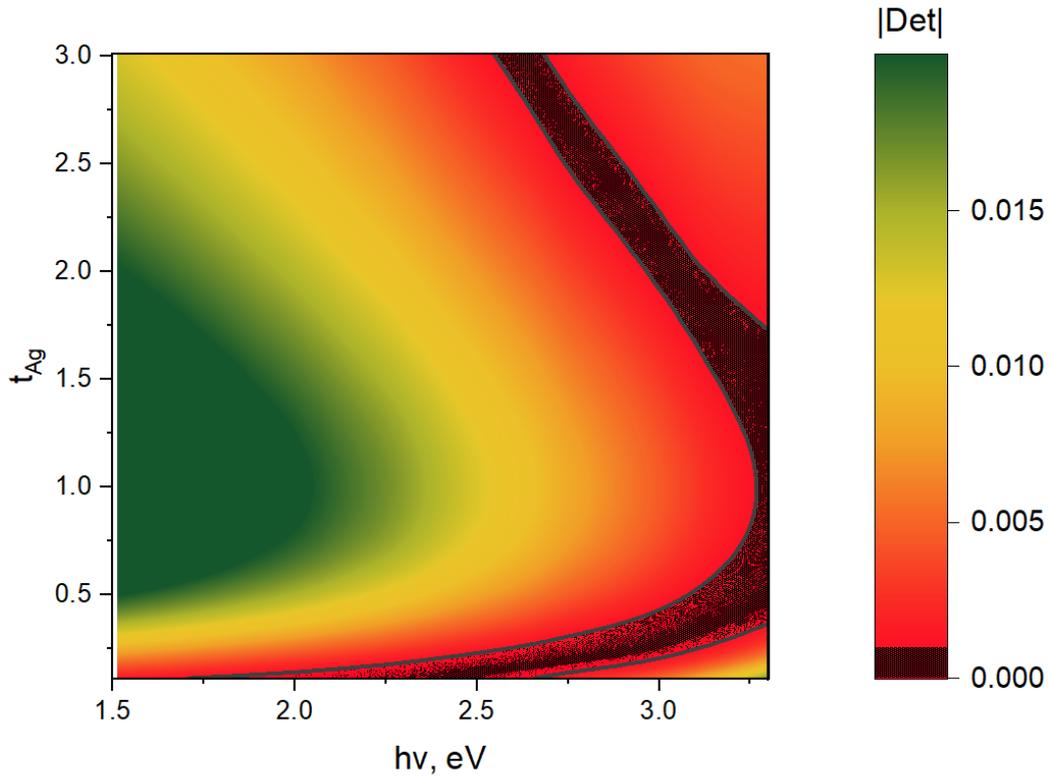

Fig. 13. The optimization map of values $M\text{det}$ calculated for a submonolayer of silver nanoparticles of diameter 20 nm seeded onto a YIG film of 100 nm thickness. The angle of incident light was considered to be 10°. The domains $M\text{det} < 0.001$ are marked by the dashed lines.

Similar calculations for the silver nanoparticles in a case of polarized light are summarized in Fig. 14. The comparison of Fig. 12 and Fig. 14, reveals that the metal, from which the nanoparticles are fabricated, should always be taken into consideration. In particular, silver can provide a more effective enhancement, when the submonolayer cover of magnetic film is formed by the silver nanoparticles.

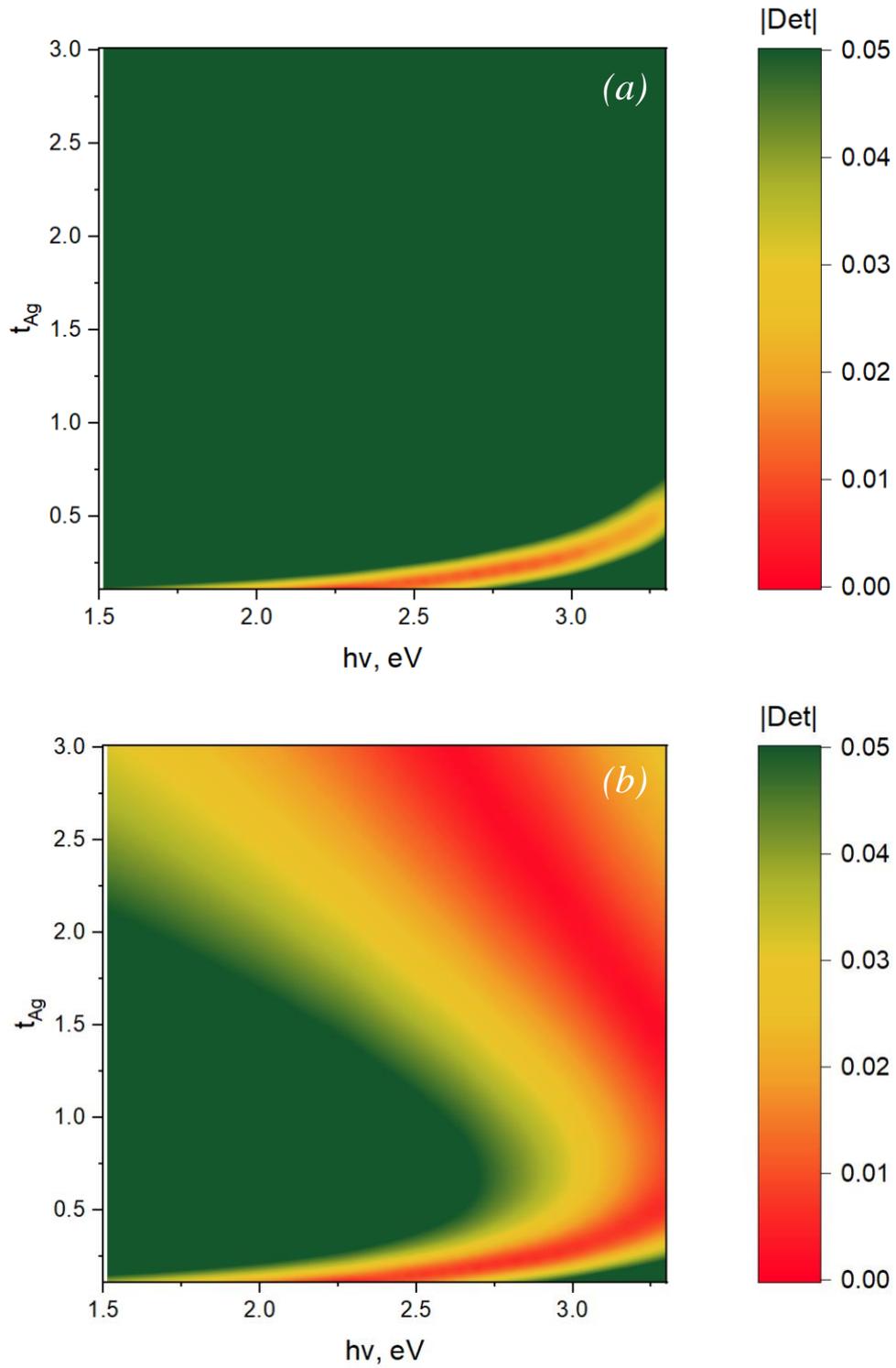

Fig. 14. The optimization map of values Mdet calculated for a submonolayer of the silver nanoparticles of diameter 20 nm seeded onto a YIG film of 100 nm thickness. The angle of incident light was set to 10°.

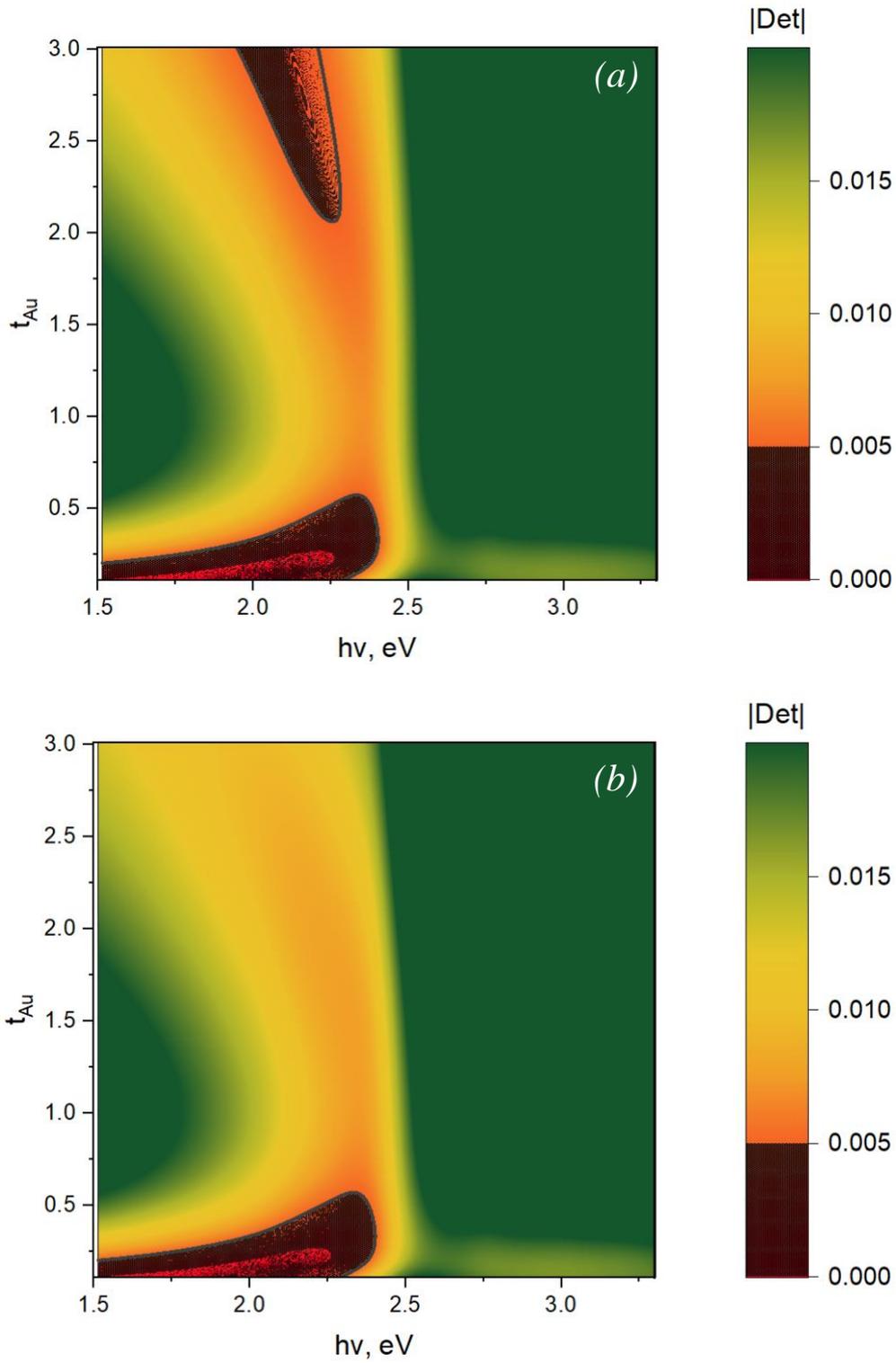

Fig. 15. The optimization map of values $M$det calculated for a submonolayer of golden nanoparticles of diameter 20 nm seeded onto a YIG film of 100 nm thickness. The angle of incident light is set to $30°$(a) and $60^0$ (b). The domains $M\det < 0.001$ are marked by the dashed lines.

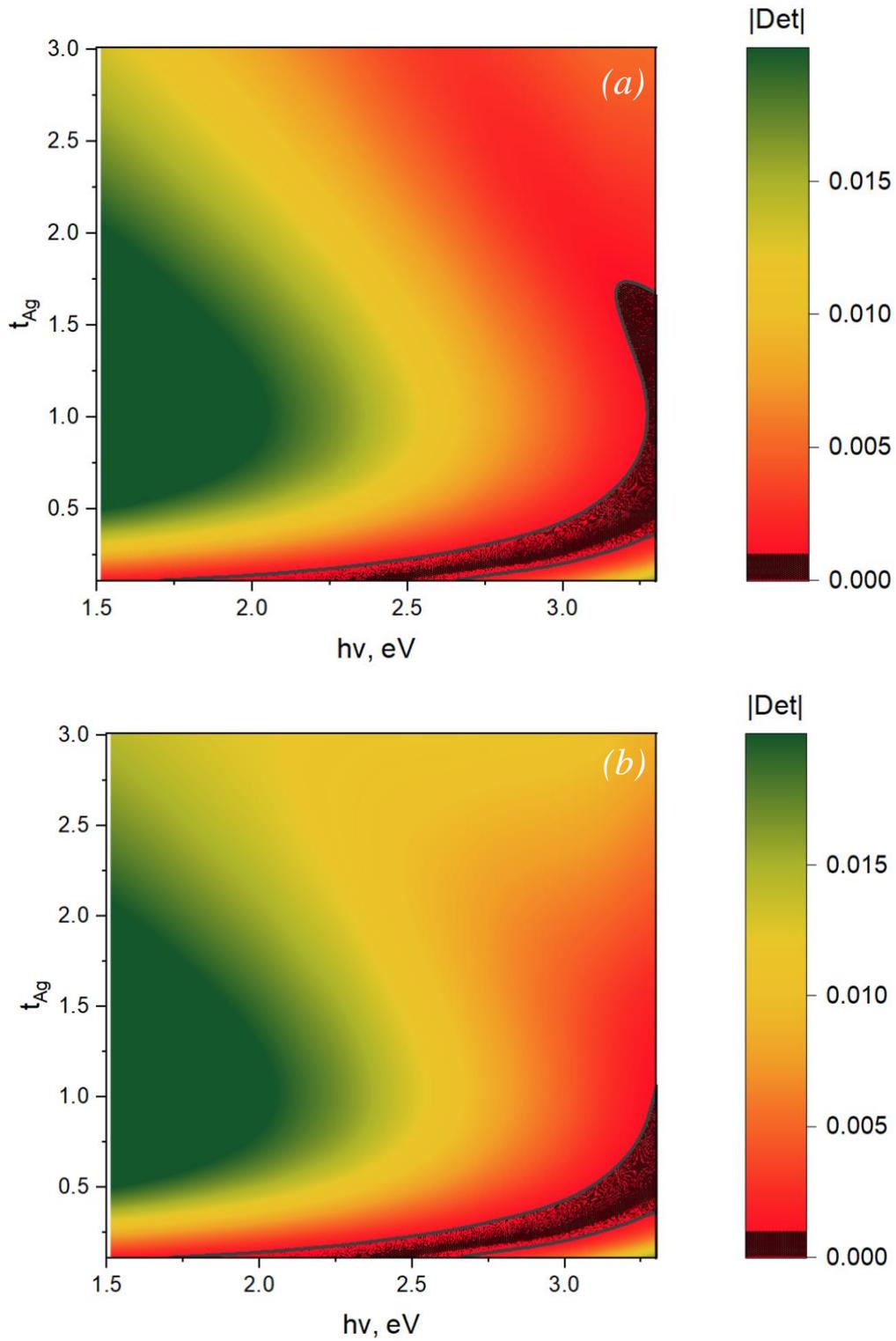

Fig. 16. The optimization map of values $M$det calculated for a submonolayer of the silver nanoparticles of diameter 20 nm seeded onto a YIG film of 100 nm thickness. The angle of incident light is 30°(a) and 60⁰ (b). The domains $M\det < 0.001$ are marked by the dashed lines.

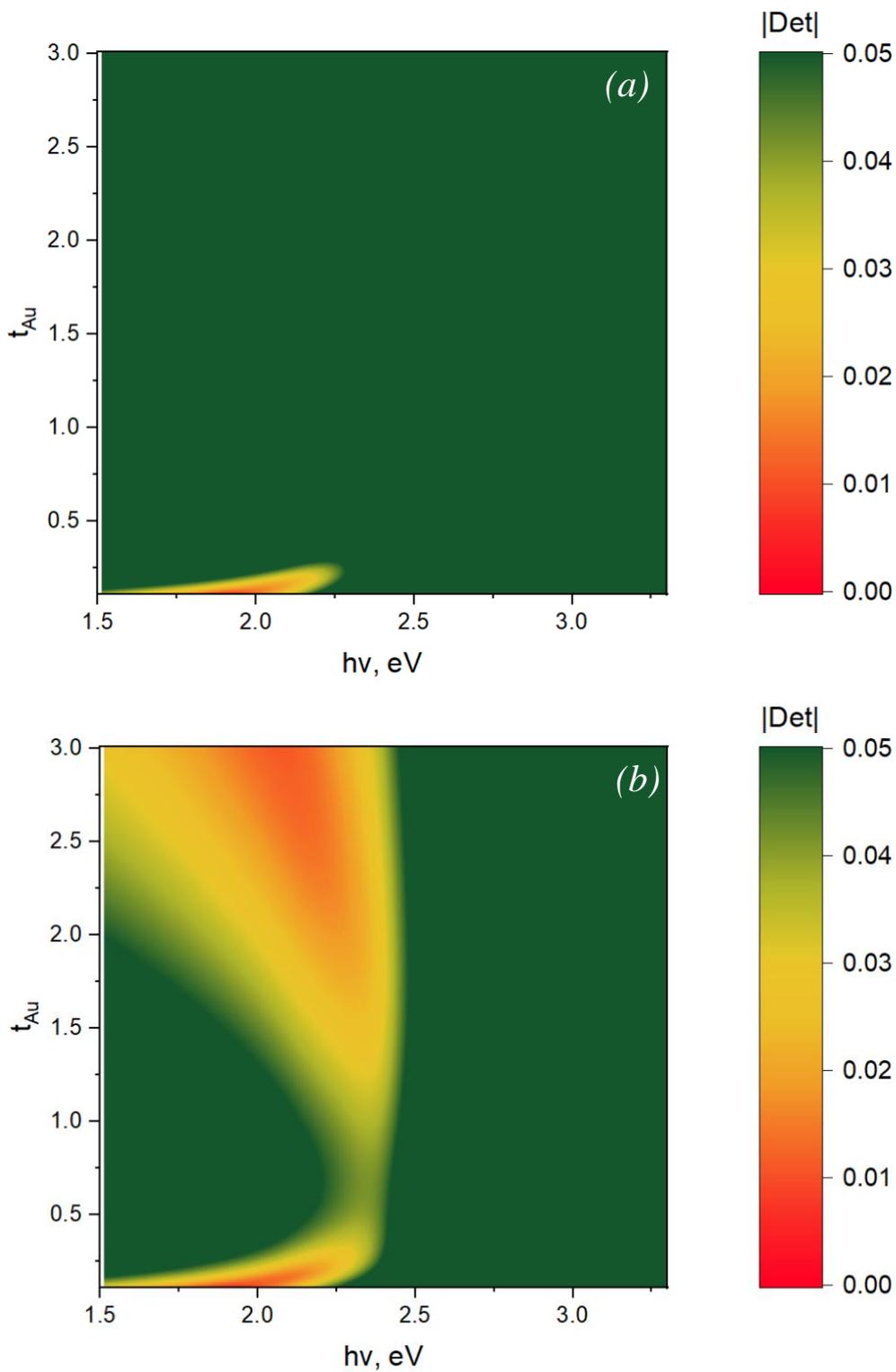

Fig. 17. The optimization map of values $M$det calculated for a submonolayer of the golden nanoparticles of diameter 20 nm seeded onto a YIG film of 100 nm thickness. The angle of incident light is 30°. The incident light is s-polarized *(a)* and p-polarized *(b)*.

The dependence of $M$det on the incidence angle $\vartheta$ is nonmonotonic. The increase $\vartheta$ up to ~ $30^0$ leads to a decrease of $M$det. Further increase of $\vartheta$ first prompts the deceleration of the decrease, and then to a slight increase of $M$det. Thus, one should take into account this nonmonotonic behavior of the effective susceptibility in order to properly choose the parameters of the experiment for the maximum enhancement of BLS signal.

The optimization maps of values $M$det calculated for a submonolayer of the golden nanoparticles of diameter 20 nm seeded onto a YIG film of 100 nm thickness are shown in Fig. 15. The angle of the incident light is set to 30° *(a)* and $60^0$ *(b)* as an example. Note, the increase of $\vartheta$ from $30^0$ to $60^0$ does not cause a decline of the interparticle interaction, which is reflected in the domain of $M$det for the small values of $t_{Au} < 0.25$ at $\hbar\omega > 2.5$ eV.

Similar calculations were performed for $M$det of a submonolayer of the silver nanoparticles (Fig. 16). Comparing a system of the magnetic film covered with golden nanoparticles, in the case of silver nanoparticles, for all possible particle shapes in the wavelength range of 500-740 nm, the dependence of $M$det on the angle of incident light is not observed. If the incident light is polarized, the optimization maps look different. The strengthening effect is somewhat weaker. Although there are still areas of nanoparticle shapes and wavelengths where BLS improvement is possible.

At the same time, the case of p-polarized light provides wider opportunities for choosing the conditions for the BRL signal enhancement. As an example, Fig. 17 presents maps of the optimization parameters map for BLS signal enhancement with a layer of golden nanoparticles considering s- and p- polarization of the incident light.

**Conclusions**

The numerical calculations based on the general approach of the theory of plasmon-assisted mechanism of the enhancement of the Brillouin Light Scattering (BLS) signal from a magnetic film were performed in this part of the work. The optimization procedure of the morphology of the nanoplasmonic structure at the surface of the magnetic film was proposed. Two kinds of nanoplasmonic structures – a single nanoobject and a submonolayer nanoparticles cover were considered.

Presented calculations suggest that the implementation of the nanoplasmonic structures for BLS signal enhancement can be more effective in the case of single nanoobjects compared to the magnetic film surface coverage with nanoparticles. The reason for this difference originates from the fact that in the case of a single nanoobject, one can achieve the local plasmon resonance (localized at the nanoplasmonic object), but not the surface plasmon resonance in the cover of magnetic film with nanoparticles. This explains why a surface wave propagation in the submonolayer of nanoparticles on top of the magnetic film surface simultaneous with radiation of BLS wave is not feasible. Indeed, in the latter case, the surface plasmon must be an evanescent wave which contradicts the necessary condition to have the scattered field simultaneously as a radiated one. It is conceivable that additional efforts may enable the simultaneous resonant excitation of surface plasmons and the radiation of Brillouin Light Scattering (BLS) light. This achievement can occur when the surface structure facilitates the transfer of the wave vectors from the radiation region to the region of eigenmodes. Such a structure can be realized based on a one-dimensional photon crystal formed by periodically placed nanoparticles along the axis of magnon propagation.

The authors anticipate that the results of the calculations will prove beneficial in guiding the enhancement of the Brillouin Light Scattering (BLS) signal and will facilitate more efficient experiments. Namely, the choice of material of the nanoparticles and spacer, their shape and concentration on the surface of the magnetic film should be considered in order to improve the BLS signal.

**Acknowledgments**

V.L. thanks the Austrian Academy of Sciences' Joint Excellence in Science and Humanities (JESH) and the ESI Special Research Fellowship for Ukrainian Scientists for the support of this work. Yu.D. and V.L. express gratitude to the IEEE Magnetics Society "Magnetism for Ukraine 2023" program. Yu.D. acknowledges support from Wolfgang Pauli Institute (WPI) Vienna. A.V.C. acknowledges the Austrian Science

Fund FWF for the support by the project I-6568 "Paramagnonics". K.O.L. acknowledges the Austrian Science Fund FWF for the support 267 through ESPRIT Fellowship Grant ESP 526-N "TopMag". The authors thank all the brave defenders of Ukraine, who made possible the finalization of the publication.

16. D. Pohl, Scanning near-field optical microscopy (SNOM), *Advances in Optical and Electron Microscopy* **12**, 243 (1991).

17. J. A. J. Backs, S. Sederberg, and A. Y. Elezzabi, A nanoplasmonic probe for near-field imaging, *Optics Express* **19**, 11280 (2011).

18. F. S. Hussain, N. Q. Abro, N. Ahmed, S. Q Memon, N. Memon, Nano-antivirals: A comprehensive review, *Front. Nanotechnol.* **4** (2022).

19. M. Chaika, S. Zahorodnya, K. Naumenko, V. Lozovski, N. Rusinchuk, Virus deformation or destruction: Size-dependence of antiviral and virucidal activities of gold nanoparticles, *ANSN* **13**, 035008 (2022).

20. V. Z Lozovski, V. S Lysenko and N. M Rusinchuk, Near-field interaction explains features of antiviral action of non-functionalized nanoparticles, *Adv. Nat. Sci.: Nanosci. Nanotechnol.* **11**, 015014 (2020).

21. H. Yu, Y. Peng, Y. Yang, et al., Plasmon-enhanced light–matter interactions and applications, *npj Comput Mater* **5**, 45 (2019).

22. V. Lozovski and A. V. Chumak, Plasmon-enhanced Brillouin Light Scattering spectroscopy for magnetic systems. I. Theoretical Model, *arXiv* (2024).

23. V. Lozovski, *J. Comput. Theor. Nanosci*. **7**, 2077 (2010).

24. Q. Wang, R. Verba, B. Heinz, M. Schneider, O. Wojewoda, K. Davídková, K. Levchenko, C. Dubs, N. J. Mauser, M. Urbánek, P. Pirro, A. V. Chumak, Deeply nonlinear excitation of self-normalised exchange spin waves, *arXiv*: 2207.0112 (2024).

25. V. Z. Lozovski, C. Lienau, G. G. Tarasov, T. A. Vasyliev, Z. Ya. Zhuchenko, Configurational resonances in absorption of metal nanoparticles seeded onto a semiconductor surface, *Physics* **12** 1197 (2019).

26. O. Keller, Local fields in the electrodynamics of mesoscopic media, *Phys. Rep.,* **268**, 85 (1996).

27. M. L. Bah, A. Akjouj and L. Dobrzynski, Response functions in layered dielectric media*, Surf. Sci. Rep*. **16**, 95 (1992).

28. A. B. Evliukhin, Sergey I. Bozhevolnyi, Surface plasmon polariton scattering by small ellipsoid particles, *Surface Science* **590,** 173 (2005).

29. I. Iezhokin, O. Keller, and V. Lozovski, Induced Light Emission from Quantum Dots: The Directional Near-Field Pattern, *J. Comput. Theor. Nanosci.* **7**, 281 (2010).29